\documentclass[fleqn,10pt,]{wlscirep}

\usepackage{subfigure}% for two column figures
\usepackage{color}

\title{Superradiant Quantum Heat Engine}

\author[1,*]{Ali \"{U}.~C.~Hardal}
\author[1]{\"{O}zg\"{u}r E. M\"{u}stecapl{\i}o\u{g}lu}
\affil{Department of Physics, Ko\c{c} University, \.{I}stanbul,Sar{\i}yer 34450, Turkey}

\affil[*]{ahardal@ku.edu.tr}

\begin{abstract}

Quantum physics revolutionized classical disciplines of mechanics, statistical physics, and electrodynamics. 
One branch of scientific knowledge 
however seems untouched: thermodynamics. Major motivation behind thermodynamics is to develop efficient heat engines. 
Technology has a trend to miniaturize engines, reaching to quantum regimes. Development of 
quantum heat engines (QHEs) requires emerging field of quantum thermodynamics. 
Studies of QHEs debate whether quantum coherence can be used as a resource. We explore an alternative 
where it can function as an effective catalyst. We propose a QHE which consists of a photon gas inside an optical 
cavity as the working fluid and quantum coherent atomic clusters as the fuel. Utilizing the superradiance, where a 
cluster can radiate quadratically faster than a single atom, we show that the work output becomes 
proportional to the square of the number of the atoms. In addition to practical value of cranking up QHE, our 
result is a fundamental difference of a quantum fuel from its classical counterpart.

\end{abstract}

\begin{document}
\flushbottom
\maketitle

\thispagestyle{empty}

%%%%%%%%%%%%%%%%%%%%%%%%%%%%%%%%%%%%%%%%%%%%%%%%%%%%%%%%%%%%%
%\section*{Introduction}
%%%%%%%%%%%%%%%%%%%%%%%%%%%%%%%%%%%%%%%%%%%%%%%%%%%%%%%%%%%%%
Superradiance (SR) was originally introduced by Dicke in 1954~\cite{dicke1954coherence} as 
a cooperative emission of light from an ensemble of excited two level atoms
in a small volume relative to emission wavelength. The atoms radiate in a
synchronized (coherent) manner, at quadratic rate
with the number of atoms. Experimental verification of SR has been accomplished in diverse systems~\cite{skribanowitz1973observation,gross1976observation,scheibner2007superradiance,
rohlsberger2010collective,devoe1996observation,eschner2001light,mlynek2014observation}. Due to fast
dephasing, typical observations are limited to pulsed or transient regimes, while modern experiments
achieve SR in steady state as well~\cite{baumann_dicke_2010}. 
Nature herself benefits from the SR processes; in particular, in  
light harvesting complexes, for 
efficient collection and transfer of solar energy~\cite{meier1997polarons,
celardo2012superradiance,scholes2002designing
}. We explore if we can follow the 
nature's example and use SR
to enhance work harvesting capability of a quantum heat engine (QHE). 

Typical QHEs are analogs of classical heat engines, whose working fluid or heat reservoirs are replaced by 
quantum systems~\cite{PhysRevLett.2.262,
PhysRevLett.93.140403,PhysRevE.76.031105,
PhysRevLett.112.150602,PhysRevLett.112.076803,
scully2003extracting,
scully2002extracting,rostovtsev2003extracting,scully2011quantum,scully2010quantum,altintas_rabi_2015}. 
While the recognition
of a quantum system, three level maser, as a heat engine goes back to 1959~\cite{PhysRevLett.2.262}, many 
modern studies discuss effects of
quantum coherence on the performance of QHEs. 
A particularly intriguing QHE consists of an optical cavity pumped by a beam
of atoms prepared in coherent superposition states.
After repeated passages of the atoms one by one~\cite{scully2002extracting}, or two by two~\cite{li2014quantum}, 
the cavity field can be thermalized to an effective temperature which can be
higher than that of the atom. 
Using such a quantum coherent thermalization procedure in a Carnot cycle, a QHE with unique properties
is proposed~\cite{scully2002extracting,scully2003extracting,rostovtsev2003extracting}. 
Photonic Carnot engine could operate with a single thermal reservoir, and could beat the Carnot efficiency. 
Addition of the coherence reservoir and the cost of coherence generation brings these feats within the boundaries 
of the second law of thermodynamics. The work output of the engine is determined by the radiation pressure on 
the cavity mirrors and it is proportional to the cavity field intensity~\cite{scully2003extracting}. 
We propose to enhance the work harvesting capabilities of a photonic QHE using SR enhancement of the cavity field 
intensity. We choose to operate our QHE
in Otto cycle instead of Carnot cycle, as it is more experimentally feasible.

We consider clusters of
$N$ two level atoms interact with the cavity at regular intervals in the ignition stage of the Otto cycle.  Steady state SR in an overdamped cluster micromaser~\cite{temnov2001superradiant,temnov2005superradiance} 
is similar to our set up, though we consider weak damping regime. The atoms are assumed to be prepared in a low 
temperature coherent superposition state and the cavity field is assumed to be in a thermal state initially. 
We find that after a number of interactions, the cavity field reaches an equilibrium state, which is a coherent thermal
state~\cite{barnett_thermofield_1985,emch_new_1986,bishop_coherent_1987,fearn_representations_1988}. 
The corresponding work output of the QHE can be determined from the
steady state photon number, which is enhanced by the SR. Accordingly it becomes proportional to $N^{2}$. 

Quantum coherence in the cluster is used here as an effective catalyst~\cite{aberg_catalytic_2014}, 
increasing the energy transfer rate from cluster to the field. It is necessary to prove that quantum coherence can be
completely recycled in order to call it as a true catalyst. As there are fundamental and practical challenges against
that~\cite{horodecki_fundamental_2013,lostaglio_quantum_2014,lostaglio_description_2015,rodriguez-rosario_thermodynamics_2013,cwiklinski2014limitations}, 
we leave it here as an intriguing open question for future contributions 
if and how quantum coherence can be recovered partially or completely. Our focus here will be to utilize quantum coherence
to enhance the rate of energy transfer. 
Coherent cluster acts like a fuel which burns quadratically faster than an incoherent one. The time to reach equilibrium
increases when the energy kicked into the cavity in rapid bursts.
This allows for harvesting quadratically more energy from a resource before equilibrium is established. Coherence
is partially  transferred from a set of clusters to the cavity field. The field can be 
described as a thermal coherent state in equilibrium~\cite{barnett_thermofield_1985,emch_new_1986,bishop_coherent_1987,fearn_representations_1988}.
In contrast to energy, which is harvested as work output, coherence is not consumed and it
remains within the field. More technically, quantum coherence is determined by the off diagonal elements of the 
density matrix of the field; and work
is harvested in a quantum adiabatic stage, where the off diagonal
elements are preserved up to a geometric and dynamic 
phase factor~\cite{berry_quantal_1984,aguiar_pinto_adiabatic_2002}. 
The diagonal elements, and hence the number of photons, 
cannot change in accordance with the adiabatic theorem. The coherence could in principle be transformed from the cavity field back 
to the clusters in the exhaust stage. The cluster
resonance frequency should be altered to match the adiabatic changes in the photon frequency. While one could envision a hybrid
system where coherence is exchanged between clusters and photons in a completely or partially reversible manner, and work is 
extracted locally from photon subsystem, there
are serious obstacles against its realization. Complete recovery of initial coherence may not be possible first due to fundamental
constraints by the time translational symmetry~\cite{lostaglio_quantum_2014,lostaglio_description_2015,rodriguez-rosario_thermodynamics_2013}; and second due to practical constraints by the rapid
dephasing of coherence. In practice one can take precautions against dephasing using time dependent control methods~\cite{erez_thermodynamic_2008}, and be
content with partial recovery of coherence within the fundamental bounds to reduce the production cost of quantum coherent fuel~\cite{scully2003extracting}. 
Instead, coherence can be repeatedly reestablished externally in the clusters before every cycle of the engine operation. This is an effective 
use of catalytic coherence. 
%
%The manuscript is organized as follows. First, we introduce our model system and the physical algorithm to achieve 
%effective thermalization. Next, we present our results for the work extraction from the quantum Otto engine. We conclude 
%with the discussions for finite temperature setups.
%%%%%%%%%%%%%%%%%%%%%%%%%%%%%%%%%%%%%%%%%%%%%%%%%%%%%%%%
\section*{The model and dynamical algorithm}\label{sec:model} 
We consider a dissipative, but high finesse, single mode optical cavity, whose field is interacting 
with a of cluster of $N$ two-level atoms for a time $t_{\mathrm{int}}$ repeatedly at a rate $r<1/t_{\mathrm{int}}$. 
A beam of clusters passing through the cavity in a time of  $t_{\mathrm{int}}$ one at a time at regular intervals 
of $1/r> t_{\mathrm{int}}$ is an equivalent alternative scenario. 

The interaction is described 
by the Tavis-Cummings Hamiltonian (in units of $\hbar=1$)
\begin{equation}\label{eq:tc}
H_\mathrm{sys}=\omega_f a^{\dagger}a+\omega_a S_z+g(aS^{+}+a^{\dagger}S^{-}),
\end{equation}
where $\omega_f$ is the cavity photon frequency, $\omega_a$ is the transition frequency of the  
atoms, and $g$ is the uniform interaction strength. The photon annihilation and creation operators
obey the boson algebra and denoted by $a$ and $a^\dag$, respectively. The atomic cluster is represented by 
collective spin operators $(S^{\pm}, S_z)=(\sum_i\sigma_i^{\pm},1/2\sum_i\sigma_i^{z})$, which
obey the $SU(2)$ spin algebra, where $\sigma_i^{\pm}$ and $\sigma_i^{z}$ are the Pauli spin matrices, corresponding to the 
transition and population inversion operators for the $i$th atom, respectively. The multi photon generalization of the Tavis-Cummings model is considered
in quantum Otto cycle from the perspective of the interplay between dynamical Stark shift, thermal entanglement
and the engine efficiency \cite{chotorlishvili2011thermal}.

The dynamics of the state of the system, $\rho_{\mathrm{sys}}$, during the interaction, is determined by the master equation
\begin{equation}\label{eq:m1}
\frac{\partial\rho_{\mathrm{sys}}}{\partial t}=-i[H_\mathrm{sys},\rho_{\mathrm{sys}}]+\frac{\kappa}{2}(2a\rho_{\mathrm{sys}}a^{\dagger}-a^{\dagger}a\rho_{\mathrm{sys}}-\rho_{\mathrm{sys}}a^{\dagger}a),
\end{equation}
where $\kappa$ is the cavity dissipation rate. 
Between the interactions, the cavity field evolves freely by the master equation
\begin{equation}\label{eq:m2}
\frac{\partial\rho_f}{\partial t}=-i[\omega_f a^{\dagger}a,\rho_{f}]+
\frac{\kappa}{2}(2a\rho_{f}a^{\dagger}-a^{\dagger}a\rho_{f}-\rho_{f}a^{\dagger}a),
\end{equation}
where $\rho_f=\mathrm{Tr}_{a}(\rho_{\mathrm{sys}})$ is the reduced density matrix of the field, 
and $\mathrm{Tr}_{a}$ is the partial trace over the atomic states. 

Every interaction starts with an initial state of the system in the form $\rho_{sys}=\rho_a(0)\otimes\rho_f$, 
where $\rho_a(0)$ is the externally prepared state of the cluster, which is the same at the beginning
of every interaction; and $\rho_f$ denote the state of the cavity field which is changing from one 
interaction to the other according to dynamics described by Eqs.~(\ref{eq:m1}) and (\ref{eq:m2}).
%%%%%%%%%%%%%%%%%%%%%%%%%%%%%%%%%%%%%%%%%%%%%%%%%%%%%%%%%
\section*{Results} \label{sec:Results}
\subsection*{Thermalization} 
%%%%%%%%%%%%%%%%%%%%%%%%%%%%%%%%%%%%%%%%%%%%%%%%%%%%%%%%%
In a usual set up, the atomic ensemble, first, gets into contact with a hot reservoir at temperature $T_h$. After a sufficient 
period of time, the state of the individual atoms $\rho_{a}^{i}$ thermalizes, i.e., 
$\rho_{a}^{i}(0)=(1/Z_a)\sum_n\exp{(-\beta E_{n}^{i})}|\psi_n\rangle^{i}\langle\psi_n|$. Here, $E_n$ and $|\psi_n\rangle$ are 
the eigenvalues and the corresponding eigenvectors of the single-atom Hamiltonian $H^{i}=(\omega_a/2)\sigma_{z}^{i}$, 
$\beta=(1/T_h)$ $(k_B=1)$ is the inverse temperature and $Z_a=\mathrm{Tr}(\rho_a^i(0))$ is the partition function. 
The state of the cluster, assuming there are no interatomic interactions, is given by $\rho_a(0)^{\prime}=\bigotimes_i\rho_a^{i}(0)$. 
The initial state of the cavity field is a thermal state at a temperature $T_c$, i.e., 
$\rho_f(T_c) = (1/Z_f)\exp{(-\beta H_f)}$ with $H_f=\omega_f a^{\dagger}a$ and $Z_f=\mathrm{Tr}(\rho_f(0))$.

We assume that $T_h$ is sufficiently low, such that most of the atoms are in their ground states. Prior to their interaction with the
cavity field, the atoms are transformed into coherent superposition states by a rotation operation
\begin{equation}\label{eq:state_a}
\rho_a(0)=R(\zeta)\rho_a(0)^{\prime}R(\zeta)^{\dagger},
\end{equation}
where $R(\zeta)=\exp{(\zeta S^{+}-\zeta^* S^{-})}$ with $\zeta=(\phi/2)e^{i\varphi}$. In our calculations we set $\varphi=0$ and take $\phi=-\pi/2$, i.e., the atomic states are moved from the pole of the Bloch sphere to the equator. The initial state of the system is given by $\rho_{sys} = \rho_a(0)\otimes\rho_f$. Our choice of the initial state is not arbitrary. It is well known that the collective atomic coherent states are closely related to the Dicke states and superradiant with the choices of $\varphi=0$ and $\phi=\pm\pi/2$~\cite{arecchi1972atomic}.
The initial state of the atomic cluster $\rho_a(0)$ can be called as a thermal coherent spin state.

\begin{figure}[!t]
\centering
     \includegraphics[width=14cm]{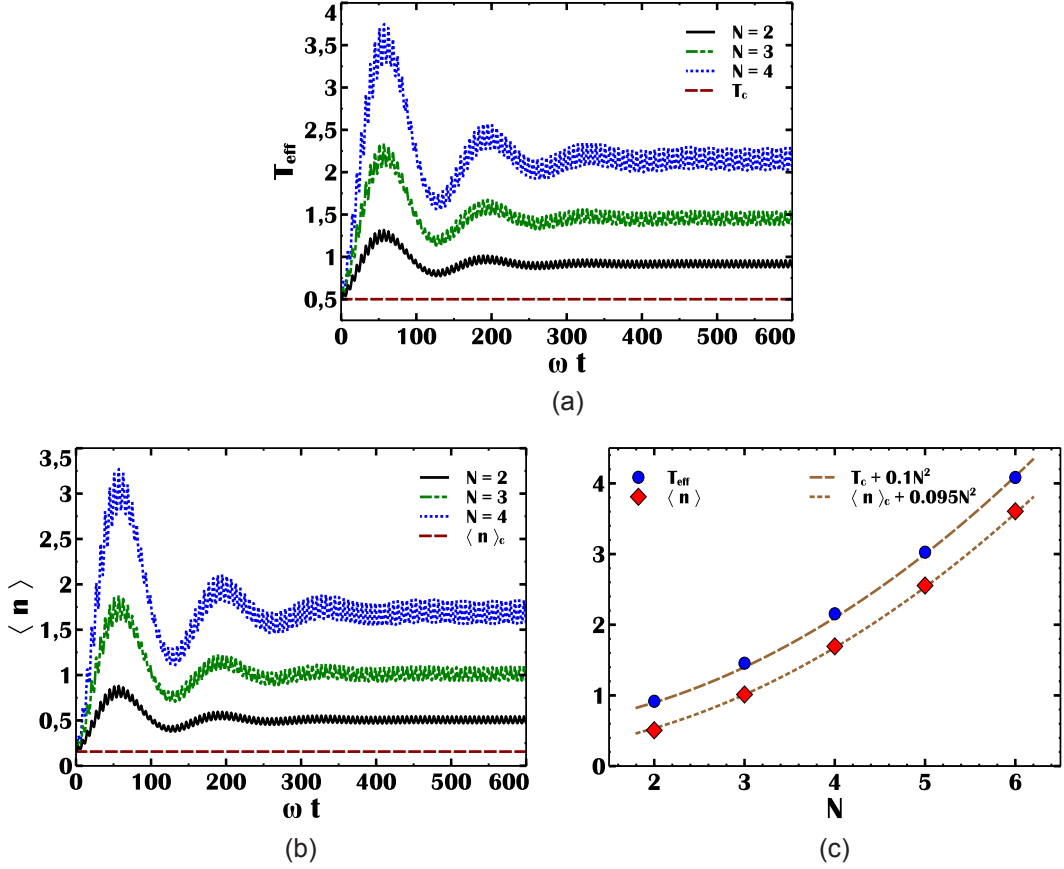}
     
         \caption{%
        \textbf{Dynamics of the effective temperature and the mean photon number.} (a) The effective temperature 
$T_{ \mathrm{eff}}$ and (b) the corresponding average number of photons $\langle n\rangle$ as a function
        of the scaled time $\omega t$, respectively, for $100$ successive interactions of $N=2$ (black-solid), $N=3$ 
(green-dashed) and $N=4$ (blue-dotted) atom clusters. Brown-dashed lines represent the initial values $T_c=0.5$ and 
$\langle n_c\rangle = 0.156$. (c) The steady state values of $T_{ \mathrm{eff}}$ and $\langle n\rangle$ as a function of 
$N$ obtained by time averaging over $250$ successive injections. All the other parameters are as explained in the text.
     }%
   \label{fig:fig2}
\end{figure}

We assume that the field and the atoms are in resonance, i.e., $\omega_f=\omega_a$, which will be used to scale energy, time and 
temperature parameters in the numerical simulations. We take $g = 0.19$, $\kappa=0.03$,  $T_h=0.001$, $T_c=0.5$
and $rt_{\mathrm{int}}=1/6$.
Since $g/\kappa>1$, our system is not in the overdamped micromaser regime and we cannot use the corresponding 
superradiant micromaser master equation~\cite{temnov2001superradiant,temnov2005superradiance}. We solve Eqs.~(\ref{eq:m1}) and (\ref{eq:m2}) instead to determine the cavity field density matrix after a number of interactions with the clusters of $N$ atoms. Our typical 
results are shown in Fig.~\ref{fig:fig2}.
In Fig.~\ref{fig:fig2}(a) and~\ref{fig:fig2}(b), we present the time dependence of the effective temperature $T_{\mathrm{eff}}$ and 
the mean number of photons in the cavity $\langle n\rangle= \mathrm{Tr}(\rho_f a^{\dagger}a)$, respectively. The initial number of 
photons in the cavity is calculated to be $\langle n_c\rangle\sim0.156$ at 
$T_c=0.5$. The effective temperature is defined by the relation 
$\langle n\rangle=1/[\exp{(1/T_ {\mathrm{eff}})}-1]$. It can be interpreted as a temperature only in a steady state which is an approximate
thermal equilibrium state. If the coherence is large, the probability distribution of the photons in the steady state 
SR phase becomes closer to that of a coherent state, with the increasing $N$~\cite{PhysRevA.81.063827}. More rigorously,
it can be described by a thermal coherent state~\cite{barnett_thermofield_1985,emch_new_1986,bishop_coherent_1987,fearn_representations_1988}. 
It can be written as a thermal state $\rho_f(T)$ subject to Glauber displacement transformation $\rho_f(T,\alpha)=D(\alpha)\rho_f(T)D^\dag(\alpha)$, with the displacement operators $D(\alpha)$, where $\alpha$
is the coherence parameter.

The results are shown for the cases of clusters with $N=2,3,4$ atoms. In each case the time range allows for
$100$ interactions between the clusters and the cavity field. We see that the cavity field takes more time to reach steady state with 
larger clusters. This observation is in fact ensures that catalytic use of coherence is consistent with the energy conservation. Each cluster 
stores an energy of $N\omega_a/2$, which scales linearly with the number of atoms. If $M$ clusters are needed to bring the cavity
field to equilibrium, then the energy delivered from clusters to the cavity will be a fraction of $MN\omega_a/2$. The change in the 
energy of the 
cavity field scales quadratically with $N$. Accordingly, if $M$ would not increase with $N$, we could in principle 
transfer more energy than the stored amount by simply increasing $N$ to extreme values. Hence, the increase of $M$ with $N$
is consistent with the energy conservation.  

Fig.~\ref{fig:fig2}(c) shows the steady state values of the effective temperature of the cavity field $T_{\mathrm{eff},ss}$ and the 
mean photon number in the cavity $\langle n\rangle_{ss}$ that are calculated via time-averaging over a period of 
$t = 1500$ corresponding to $250$ successive interactions of $N=2,3,4,5,6$ atom clusters. Curve fitting yields the 
relations $T_{\mathrm{eff},ss} = 
T_c + 0.1N^2$ and $\langle n\rangle_{ss}=\langle n\rangle_c+0.095N^2$. The proportionality constant $\xi\sim0.1$, that appears in front of the $N^2$ scaling, is  due to the single atom micromaser emission intensity. In micromasers with random arrival times of pumped atoms, the analytical expression of the emission intensity is given by $I_{1}=rg^2t_{int}^2P_e/\kappa$~\cite{liao2010single} where $P_e$ is the probability of finding the two level atom in its excited state. Our parameters used in this formula indeed verifies that incoherent single atom emission is $\sim0.1$. On the other hand, we cannot immediately conclude that this formula is
applicable to the our set up of regular injection of clusters, for which case the analytical verification of such a relation is proven to be a difficult problem.
Nevertheless, we further test these relations via numerical investigations and found that
$\xi\propto4g^{2.3}$ for $\kappa=0.03$ and $\xi\propto0.008\kappa^{-0.7}$ for $g=0.19$. These numerical
fits comply with the analytical estimation.

The mean number of photons in a thermal coherent
state (TCS) is given by $\langle n\rangle_\mathrm{TCS}=\langle n\rangle_\mathrm{th}+\mid \alpha\mid^2$. The effective temperature
description is not essential to comprehend or operate the thermalization. We recognize 
that the physical temperature of the cavity remains the same and the atoms only transfer coherence to the field. Clusters therefore
act as a pure coherence reservoir. By fitting TCS to the numerically determined density matrix in equilibrium, with fidelity $\sim 1$, we verify that $\mid \alpha\mid^2\sim 0.1N^2$. This quadratic enhancement of coherence transfer 
into the cavity field can be translated into  
useful work output, by associating number of photons with the radiation pressure, in a photonic QHE.
%%%%%%%%%%%%%%%%%%%%%%%%%%%%%%%%%%%%%%%%%%%%%%%%%%%%%%%%%%%%%%
\subsection*{Superradiant quantum Otto engine}
% %%%%%%%%%%%%%%%%%%%%%%%%%%%%%%%%%%%%%%%%%%%%%%%%%%%%%%%%%%%%%%
We consider a four-stroke quantum Otto cycle for our photonic QHE. The working fluid, which is the photons inside the
cavity, is described by 
Hamiltonian $H_{f}=\omega a^{\dagger}a$ with eigenvalues $E_{n}, (n=0,1,2...)$. The corresponding 
eigenstates $\mid n\rangle$ has occupation probabilities $P_{n}(T)=\exp{(-\beta E_{n})}/Z$. 

In the ignition
stroke, the photon gas is heated in a quantum isochoric process to temperature $T_H$ and the occupation probabilities 
change to $P_n(T_{H})$.
The eigenvalues remain constant and denoted by $E_{n}^H$. There is only heat intake and no work is done.
This process is assumed to happen by the superradiant thermalization procedure where the photons interact 
with coherent atomic clusters. For simulations we choose the starting temperature of the cavity field as
$T_c=0.5$, in units of  $\omega_H$, which is the frequency of the atoms and the field. 
After the thermalization, the photon gas effective temperature becomes $T_H\equiv T_{\mathrm{eff},\mathrm{ss}}$.
Both the energy and coherence
of the clusters are partially transferred to the photon gas, which can be described approximately in thermal state
in the case of weak coherence. More generally, we do not need an effective temperature description and one can consider larger
coherence transfer by thermalization into a thermal coherent state. The clusters and the cavity field comes into an equilibrium
in terms of coherence, rather than temperature. This vision is in parallel with the recent ideas on thermodynamics
of quantum coherence~\cite{rodriguez-rosario_thermodynamics_2013,lostaglio_quantum_2014,lostaglio_description_2015,cwiklinski2014limitations}
as well as with other generalizations of second law and thermodynamical principles~\cite{aberg_truly_2013,brandao_second_2015,skrzypczyk_work_2014}. 
If we denote the initial thermal density matrix of the field
as $\rho_f(T_c)$, it evolves into a coherent thermal state $\rho_f(T_c,\alpha)$
in the steady state where $\mid \alpha\mid^2\sim 0.1N^2$.

Second stage is the expansion stroke, which is a quantum adiabatic process, 
where the work is done by the photon gas and the eigenvalues change to $E_n^L$ by changing the frequency to 
$\omega_L$; while there is no heat
exchange and the occupation probabilities remain the same, the physical temperature drops from $T_c$ 
by expansion. The off diagonal elements remains the same as well up
to a dynamical and a geometrical phase factor. The 
diagonal elements of the density matrix, and hence the photon number, 
remains the same and the magnitude of the coherence is preserved. Ideally the coherence would not vanish in the 
adiabatic process, but in practice decoherence and dephasing would reduce coherence. The density matrix of the field at the end of the expansion can be written as 
$\rho_f(T^\prime,\alpha^\prime)$.

In the subsequent exhaust stroke, the state of the photon gas should be transformed into a thermal one as $\rho(T_L)$
in another quantum isochoric process and the occupation probabilities change to $P_n(T_L)$,
but the eigenvalues remain constant. An amount of heat released, but no work is 
done. Coherence can be transferred to environment by dephasing or in principle
to clusters. If the same clusters would be used, it is necessary to alter their
frequencies to $\omega_L$.
Due to conservation of the photon number and by variation of the atomic frequency, the coherence transfer is energetically allowed,
though dephasing and time translation symmetry constraints makes it limited~\cite{lostaglio_quantum_2014,lostaglio_description_2015,rodriguez-rosario_thermodynamics_2013}. 
Accordingly one could choose more practical methods than using clusters to
cool the cavity. The thermal density matrix of the field at the end of the exhaust stroke can be written as 
$\rho_f(T_L)$.

Compression stroke is the last stage where the eigenstates change back to $E_n^{H}$ by variation of the photon frequency
from $\omega_L$ to $\omega_H$ in another quantum adiabatic process where work is done on the cavity field without any
heat exchange; and the temperature raises back to $T_c$. The density matrix of the cavity field is set back to $\rho_f(T_c)$.
If the same clusters are used in cooling stage, their frequency should be changed back to $\omega_H$.
Due to dephasing and fundamental
constraints on coherence transformations~\cite{lostaglio_quantum_2014,lostaglio_description_2015,rodriguez-rosario_thermodynamics_2013}, the clusters would need to be
induced coherence at the beginning of every engine cycle, so that their state is set back to $\rho_a(0)$. The coherence production costs can be reduced by optimizing time dependent control methods to reduce dephasing~\cite{erez_thermodynamic_2008} and by using schemes to recover
coherence as much as possible. 

In our calculations, we employ the quantum mechanical interpretation of the first law of thermodynamics, where the heat absorbed $Q_{in}$, the heat released $Q_{out}$, the net work done 
$W$ are given by the relations~\cite{PhysRevLett.93.140403}
\begin{eqnarray}\label{eq:hwe}
Q_{in}&=&\sum_nE_n^H\left[P_n(T_H)-P_n(T_L)\right],\nonumber\\
Q_{out}&=&\sum_nE_n^l\left[P_n(T_L)-P_n(T_H)\right],\nonumber\\
W&=&Q_{\mathrm{in}}+Q_{\mathrm{out}}
=\sum_n\left[E_n^H-E_n^l\right]\left[P_n(T_H)-P_n(T_L)\right],
\end{eqnarray}
where $E_n^H$ ($E_n^L$) are the energy levels during the isochoric stages. 
The efficiency $\eta$ is defined by $\eta=W/Q_{\mathrm{in}}$.
Throughout our analysis, we only consider 
positive work extraction which obeys the relation $Q_{\mathrm{in}}>-Q_{\mathrm{out}}>0$ in accordance with the 
second law of thermodynamics. 

\begin{figure}[!t]
\centering
     
     \includegraphics[width=14cm]{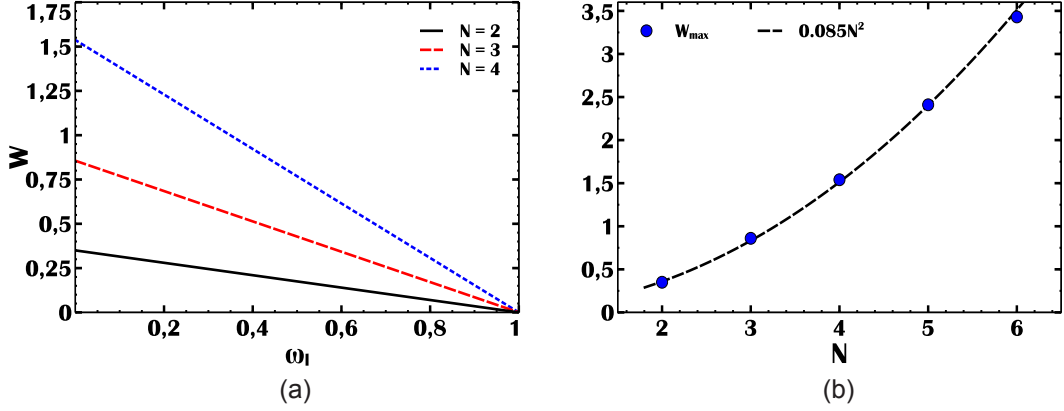}
         \caption{
        \textbf{Work output (W) of the photonic quantum Otto engine.} (a)  Work harvested by the photonic quantum 
Otto engine for $N=2$ (black-solid), $N=3$ (red-dashed) and $N=4$ (blue-dotted) atomic cluster pumps, as 
a function of  frequency $\omega_L$ (in units of $\omega_H$). (b) Maximum work output ($W_{\mathrm{max}}$, in
units of $\omega_H$) 
from the engine as a function of $N$. 
        }
   \label{fig:fig3}
\end{figure}
Using these definitions, it is straightforward to show that the work output is proportional to the 
difference in the mean number of photons~\cite{PhysRevE.76.031105}
$W=\eta(\langle n\rangle_{\mathrm{ss}}-\langle n\rangle_L)$, where $=\langle n\rangle_L=\langle n\rangle_c$ and
$\eta = 1-\omega_L$. These relations are plotted in Fig.~\ref{fig:fig3}. The work output $W$ is shown in 
Fig.~\ref{fig:fig3}(a) for atomic clusters of 
$N=2,3,4$ atoms. The positive work condition is independent of $N$ and given by $\omega_L<\omega_H$. 
In the simulations, we find the temperature of the photon gas before the compression stroke $T_L$ for each $\omega_L$, 
using the relation $T_L = \omega_L T_c$ which gives $T_L= \omega_L/2$ for $T_c=0.5$. 
This relation is a consequence of the requirement to retain same
occupation probabilities in the quantum adiabatic process.
The positive work assumes its maximum at maximum efficiency when $\omega_L\ll1$, accordingly, at low $\omega_L$,
photonic quantum Otto engine is able to translate SR enhancement of cavity intensity to its work output, as shown in 
Fig.~\ref{fig:fig3}(b). We find that the harvested work is maximum
at maximum efficiency and it obeys a power law such that $W_{\mathrm{max}}\sim0.085 N^{2}$ at $\eta_{\mathrm{max}}\sim0.99$.
%%%%%%%%%%%%%%%%%%%%%%%%%%%%%%%%%%%%%%%%%%%%%%%%%%%%%%
\section*{Discussion}
%%%%%%%%%%%%%%%%%%%%%%%%%%%%%%%%%%%%%%%%%%%%%%%%%%%%%%%
We considered a photonic QHE that undergoes a four-stroke Otto cycle. The working substance is taken as a
photon gas in an optical cavity. The coherent thermalization of the photon gas has been accomplished via 
coupling two-level atom clusters, acting as quantum coherent fuel, with the cavity at regular time intervals.
The cluster atoms have initially been prepared in a thermal coherent spin state. The cavity field reaches an
equilibrium coherent thermal state. We find that the mean number of the cavity photons 
and the corresponding effective temperature are scaled with the square of the number of the atoms in the clusters, due to SR.
We argued that the coherence of quantum fuel function as an effective catalyst to the engine cycle. It makes the 
energy transfer to the working substance faster. It is not called as true catalyst due to fundamental and practical 
constraints on its complete recovery
by the time translation symmetry as well as rapid dephasing. Instead, it can be restored before every interaction externally to
imitate an effective catalytic use of coherence for enhancing energy transfer rate. 
SR enhanced thermalization is translated to work output of the photonic Otto engine, so that it is scaled quadratically with
the number of atoms in the cluster. Such a scaling law is a quantum coherent effect, which cannot be realized with classical fuels, 
and hence exhibiting a profound difference of quantum fuels from their classical counterparts. 

Our proposed engine could work at higher temperatures, though still less than atomic and photon frequencies. In such
a case our model should be generalized to Dicke Hamiltonian which includes counter rotating terms.
Instead of preparing clusters at low temperatures, they could be at higher temperatures, including $T_h>T_c$.
Beyond a critical coupling strength and below a critical temperature, such a model predicts a superradiant phase. 
However, the critical coupling is in the ultrastrong interaction regime, $g/\omega>1$~\cite{wang1973phase},
and challenging experimentally. Instead, if we keep inducing coherence externally to the clusters, we find SR could
still occur at a lower coupling strengths. Specifically, at $g/\omega\sim0.36$, which is in SR regime, numerical simulations
give $T_{\mathrm{eff},\mathrm{ss}}=T_c+N^2$ at finite $T_a,T_c<1$. The fluctuations around steady state values are less in the Dicke model case. 
Another advantage of inducing coherence is that the SR emission relies less on the critical time $t_{\mathrm{crt}}$~\cite{andreev1980collective}, which is required for the atoms to build up collective coherence inside the cavity.
It is inversely proportional to the injection rate of the atoms in non-dissipative systems~\cite{andreev1980collective}.  By externally
inducing coherence, we do not need to increase the injection rate to initiate SR earlier. Combination of noise induced coherence 
schemes~\cite{scully2011quantum} with the coherence reservoir can be an attractive extension of superradiant QHE from the 
perspective of cost of coherence generation. 

In contrast to infinitely slow photonic Carnot engine, superradiant QHE in Otto cycle can produce finite power. 
It is necessary to make a dynamical simulation of the full cycle to examine the power output. We can predict the power scales
linearly with the number of atoms in the cluster, as the thermalization time increases with the cluster population. 

We have recently examined a single multilevel atom as a quantum fuel for a photonic Carnot engine~\cite{turkpencce2015quantum}. 
The atom was assumed to have multiple ground
states in coherent superposition, which are coupled to a common excited state by the cavity field. This effective model allows
for analytical determination of the steady state photon number and related thermodynamical properties in Carnot cycle. It is 
found that the work output is scaled quadratically with the number of quantum coherent ground states. Rapid dephasing
of the multilevel coherence as well as effective nature of the proposed level scheme and the vanishing power of the Carnot cycle
make the benefits of quantum coherence in a single multilevel atom as a quantum fuel severely limited.  
An intriguing question is then if the quadratic enhancement in work harvesting capability, obtained both 
for a single multilevel atom as well as for many atoms as quantum fuels, is a fundamental limiting power law associated
with the number of quantum resources; or 
if there can be more advantageous power laws, which can be translated into power output as well. Indeed,
one can imagine combination of multilevel and SR enhancements. 
There are proposals that quantum entangled initial preparations for SR from multilevel systems could give
faster than quadratic increase in intensity ~\cite{agarwal2011quantum}. Alternatively, SR emission in
photonic crystals can also yield higher power laws~\cite{john_localization_1995}. 
Our superradiant photonic QHE could be further improved by 
such extensions. 

Atomic clusters with effective catalytic quantum coherence can be used in other QHE cycles and systems.
Their enhanced energy transfer rates make them fundamentally distinct quantum fuels from classical ones, 
and allow for technologically appealing power laws in the work output of QHEs.
%%%%%%%%%%%%%
\section*{Methods}\label{sec:Methods}
\subsection*{Computational algoritm}
We run our simulations by using scientific python packages and some key libraries of QuTiP python~\cite{johansson2012qutip}. The numerical algorithm is prepared as one to one correspondence with the algorithm of the physical model. Inability to make adiabatic elimination of field mode ($g^2N/\kappa^2\gg1$) and the requirement to take partial traces over
atomic degree of freedom after each injection as well as the exponential increase of the Hilbert space dimension with the addition of atoms to the cluster set limits to the maximum number of atoms to be used to thermalize the working fluid. Again, the requirement to take partial traces over atomic degree of freedom makes the quantum Monte-Carlo trajectory methods unsuitable and force us to use full master equation approach. 

Moreover, the requirement of ordered injection of atomic clusters makes parallel programming over multi-cores almost impossible. On the other hand, we use multi-threading (parallel computing over a single core) to accelerate the code and reduce the run-time, though parallelising has no effect on the required dimension of the Hilbert space.
\subsection*{Effects of atomic decoherence}
The decoherence of the atomic clusters contributes to the dynamics as the spontaneous emission and the dephasing. The most general Lindblad dissipators reads:
\begin{eqnarray}\label{eq:at_dec}
\nonumber L(x)\rho &=& \frac{\gamma}{2}(2x\rho x^{\dagger}-x^{\dagger}x\rho-\rho x^{\dagger}x),\\
L(y_i)\rho &=& \frac{\gamma}{2}\sum_{i=1}^{N} (2y_i\rho y_i^{\dagger}-y_i^{\dagger}y_i\rho-\rho y_i^{\dagger}y_i)
\end{eqnarray} 
where $x\in \{S^{-}, S_{z}\}$ if the contributions are collective and $y_i\in \{\sigma_i^{-}, \sigma_{z}^{i}\}$ if the contributions are individual with a dissipation coefficient of $\gamma$. 
Our simulations revealed that the superradiance emission profile prevails with slightly smaller effective temperatures
for decoherence rates changing from $0$ to $6\kappa$ range and has the same behaviour as in in In Fig.~\ref{fig:fig2}(a) and~\ref{fig:fig2}(b). This result follows from the fact that the passage times of the
atomic clusters are small for the atomic decoherence to kick in and the transfer rate of the quantum coherence is robust ($\sim N^2$).
 The most detrimental effect is found to be pure dephasing and
hence we consider pure dephasing rate as the main limiting factor
to determine a most suitable experimental set up.

In a typical microwave resonator with a frequency of $\omega\sim51$~GHz, the atomic dephasing time $T_2\sim116$~$\mu$s leads to the dephasing rate $\gamma_{\phi}\sim10^4$~Hz~\cite{blais2004cavity}. The magnitude of the dissipation is about $\kappa\sim10^3$~Hz~\cite{blais2004cavity}, therefore $\gamma_{\phi}\sim10\kappa$. Instead, if we consider an optical resonator system with a frequency of $\omega\sim350$~THz, we have $\gamma_{\phi}\sim\kappa\sim10^8$~Hz, though the coupling strength is about $g\sim220$~MHz and therefore $g\sim\kappa$~\cite{blais2004cavity}. Thus, in microwave and optical resonator systems, the superradiance conditions are not satisfied. A circuit quantum electrodynamics set up, on the other hand, with $\omega\sim10$~GHz, $g\sim100$~MHz, $\gamma_{\phi}/\omega\sim5\times10^{-6}$ and $\kappa/\omega\sim6\times10^{-4}$~\cite{blais2004cavity} seem to be most suitable modern resonator set up to satisfy the conditions for superradiant heat engine.

Using circuit QED parameters, our numerical simulations give a range of $30-60$~ns thermalization time for the number of atoms that we considered in the manuscript. The interaction time is taken to be $\sim0.1$~ns. An optimization of the interaction time relative to atomic decoherence times require further examination to test if available control methods on atom-field coupling and uncoupling in circuit QED can be sufficiently fast. In addition, thermalization time is much larger than the typical time of adiabatic stages which should be larger than $1/\omega\sim0.1$~ns. The power output of the engine is therefore limited by the thermalization time. 
\subsection*{Cost of coherence}
Let us consider a two level atom which is in its ground state $| g\rangle$. To create a coherent superposition state $|\psi\rangle=(1/\sqrt{2})(| g\rangle+| e\rangle)$, we may
apply a square pulse with pulse duration $\tau_p$ and pulse amplitude $E_p$. The area of the pulse is given by the relation $A_p = dE_p\tau_p/\hbar$ with $d$ being the magnitude of the dipole-matrix element such that the Rabi frequency is given by $\Omega_R=dE_p/\hbar$. We can approximate the pulse area with the tipping angle in the Bloch sphere, i.e, $A\sim\theta$ which is necessarly equal to $\pi/2$ to obtain a coherent superposition state. If we denote the spontaneous emission rate of the two level atom with $\gamma$, Fermi's golden rule leads ($n\approx1$)
\begin{equation}
d^2=\frac{3\pi\epsilon_0\hbar c^3\gamma}{\omega},
\end{equation}
where $\epsilon_0$ is the vacuum permittivity, $c$ is the speed of light and $\omega$ is the resonant frequency of the field. We, then, obtain for the pulse amplitude that
\begin{equation}
E_p=\frac{\hbar\pi}{2\tau_p}\sqrt{\frac{\omega^3}{3\pi\epsilon_0\hbar c^3\gamma}},
\end{equation}
and thus, the pulse intensity reads
\begin{equation}
I_p=\frac{c\epsilon_0}{2}|E_p|^2=\frac{\pi\hbar\omega^3}{24c^2\tau_p^2\gamma}.
\end{equation}
For a beam with a width of $\delta$, the pulse energy is given by
\begin{equation}
U_p=I_p\pi(\delta/2)^2\tau_p=\hbar\omega\frac{\pi^2\omega^2\delta^2}{96c^2\tau_p\gamma}.
\end{equation}
We make use of the relations $\omega=2\pi c/\lambda$ and $\zeta=\lambda/\pi\delta$, with $\lambda$ and $\zeta$ being the wavelength of field and the radial beam divergence, respectively. We obtain
\begin{equation}
U_p=\hbar\omega\frac{\pi^2}{24}\frac{1}{\tau_p\gamma}\frac{1}{\zeta^2}\sim\frac{\pi^2}{3}\hbar\omega\sim3\hbar\omega,
\end{equation}
where we set $1/\tau_p\gamma\sim2$ and $\zeta\sim0.5$. For an $N$ atom cluster, the cost of coherence is found to be $U_{\text{cost}}^{\prime}=NU_p$. Thus, if we need $m$ clusters to obtain superradiant themalization of the working fluid, the total cost of coherence is given by $U_{\text{cost}}=mU_{\text{cost}}^{\prime}$. 

If we use $N=2$ atom ensembles $m=250$ times to thermalize the working fluid, the maximum work output obtained from the superradiant heat engine is found to be $W_{\text{out}}\sim0.35\omega\sim10^{-25}$J for a $\omega/2\pi=5$~GHz resonator. This output is, then, at least three order of magnitude smaller than the cost $U_{\text{cost}}=1500\omega\sim5\times10^3 W_{\text{out}}$.

It is crucial to distinguish the cost paid to generate the coherent atoms from the cost to maintain coherence in the working substance. The housekeeping cost to maintain coherence in working fluid can reduce the thermodynamic efficiency~\cite{gardas2015thermodynamic}. In our case the atoms are the fuel and their coherence is the catalyst. Photon gas in the cavity is the working substance. Coherence does not need to be maintained in the working substance, photon gas, in the operation cycle of the engine. The generation costs associated with in the fuel are not included to the thermodynamical efficiency of engines. Accordingly we do not include the cost of coherence to the engine efficiency determination.

Our calculation of cost of coherence produces a result larger than the work output of the engine and verifies thermodynamical constraints. One could consider a definition of round trip efficiency as another figure of merit for the relative energy harvested with respect to the total energy spent for the engine. From this perspective quantum coherence generation cost must be reduced to make the proposed engine more appealing for certain applications.
%%%%%%%%%%%%%%%%%%%%%%%%%%%%%%%%%%%%%%%%%%%%%%%%%%%%%%%%%%%%%%%%%%%%%%%%%

%%%%%%%%%%%%%%%%%%%%%%%%%%%%%%%%%%%%%%%%%%%%%%%%%%%%%%%%%%%%%%%%%%%%%%%%%
%
\section*{Acknowledgments}
We thank G.~S.~Agarwal, A.~Imamoglu, H.~Tureci, I.~Adagideli and J.~Vaccaro for illuminating discussions. 
A.~\"{U}.~C.~H.~acknowledges the COST Action 
MP1209. A.~\"{U}.~C.~H.~and \"{O}.~E.~M.~gratefully 
acknowledge hospitality of Princeton University Electrical Engineering Department
where early phases of this work are developed.
A.~\"{U}.~C.~H.~and \"{O}.~E.~M.~acknowledge the support 
from Ko\c{c} University and Lockheed Martin Corporation Research Agreement.   .

\section*{Author contributions statement}
\"{O}.~E.~M.~conceived the idea and developed the theory. A.~\"{U}.~C.~H.~derived the technical results and 
carried out numerical simulations. A.~\"{U}.~C.~H.~and \"{O}.~E.~M.~wrote the manuscript.

\section*{Additional information}
Competing financial interests: The authors declare no competing financial interests.

\end{document}